\documentclass[10pt,twocolumn,letterpaper]{article}

\usepackage{cvpr}              

\usepackage{graphicx}
\usepackage{amsmath}
\usepackage{amssymb}
\usepackage{booktabs}
\usepackage{multirow}
\usepackage{float}

\makeatletter
\@namedef{ver@everyshi.sty}{}
\makeatother
\usepackage{tikz,pgfplots}

\usepackage[pagebackref,breaklinks,colorlinks]{hyperref}

\usepackage[capitalize]{cleveref}
\crefname{section}{Sec.}{Secs.}
\Crefname{section}{Section}{Sections}
\Crefname{table}{Table}{Tables}
\crefname{table}{Tab.}{Tabs.}

\def\cvprPaperID{29} 
\def\confName{CVPR}
\def\confYear{2022}

\newcommand{\pastref}{\hat{\mathbf{x}}_{p}}      
\newcommand{\futureref}{\hat{\mathbf{x}}_{f}}    
\newcommand{\pastflow}{\mathbf{v}_{p}}           
\newcommand{\futureflow}{\mathbf{v}_{f}}         
\newcommand{\pred}{\tilde{\mathbf{x}}_t}         
\newcommand{\bbeta}{\boldsymbol{\beta}}          

\definecolor{lightblue}{HTML}{a3b4d8}
\definecolor{charcoal}{HTML}{264653}
\definecolor{persiangreen}{HTML}{2A9D8F}
\definecolor{orangeyellow}{HTML}{E9C46A}
\definecolor{sandybrown}{HTML}{F4A261}
\definecolor{burntsienna}{HTML}{E76F51}

\begin{document}

\title{Artificial Intelligence based Video Codec (AIVC) for CLIC 2022}

\author{Th\'eo Ladune, Gordon Clare, Pierrick Philippe, F\'elix Henry
\\
Orange Innovation, France\\
{\tt\small firstname.lastname@orange.com}
}
\maketitle

\begin{abstract}
  This paper presents the AIVC submission to the CLIC 2022 video track. AIVC is
  a fully-learned video codec based on conditional autoencoders. The flexibility
  of the AIVC models is leveraged to implement rate allocation and frame
  structure competition to select the optimal coding configuration per-sequence.
  This competition yields compelling compression performance, offering a rate
  reduction of $-26$~\% compared with the absence of competition.
\end{abstract}

\section{Introduction}

Over the last few years, deep neural networks (DNN) have been proposed to
replace the handcrafted operations performed by conventional video coding
algorithms (\textit{e.g.} HEVC \cite{hevc} and VVC \cite{vvc}). Thanks to their
expressiveness and their ability to be globally optimized, DNN-based coding
schemes have been showing fast-paced performance improvement. For instance, the
learned image coder proposed by Cheng \textit{et al.} \cite{cheng2020learned} is
competitive with the image coding configuration of VVC.

Yet, the introduction of an additional temporal dimension makes video coding
substantially more difficult than image coding. Even though numerous previous
works address learned video coding
\cite{guo2021learning,Agustsson_2020_CVPR,9506275,hu2020improving,hu2021fvc,ladune2022aivc},
conventional algorithms remain the best performing approaches. This assessment
is supported by the CLIC 2021 video track results, which was won by a
perceptually optimized VVC submission \cite{Philippe2021a}.
\bigskip

The processing of a video frame by a conventional and a learned video codec
share many similarities. Both approaches compute a prediction of the frame to
be coded, based on previously transmitted frames. Then only the unpredicted part
is sent. While conventional and learned codecs rely on the same basic principles,
a significant difference arises when one takes a closer look. For each step of
the processing, conventional approaches feature a large number of coding modes
(\textit{i.e.} different operations available), enabling the selection of the
most suited operation for the signal to compress. This selection is performed at
a pixel level, allowing for fine-grain content adaptation. On the contrary,
learned approaches tend to perform the same processing on all input data, which
limits content adaptation.

The goal of the CLIC 2022 video track is to compress several 10-second video
sequences into an allocated bits budget, while obtaining the best reconstruction
quality. These videos exhibit a great variety of spatial and temporal
complexity. As such, it is possible to save bits for simple signals (which
require fewer bits to be properly reconstructed) and invest these
bits into more complex content. While this competition is easily achieved for
conventional codecs, it is more challenging for learned ones.

In this paper, we propose to use AIVC \cite{ladune2022aivc}, an open-source
learned video codec, and to adapt it for the CLIC 2022 video track. In order to
cope with the wide variety of the video sequences, we carry out large-scale
competition among the different possibilities offered by AIVC:~multiple quality
levels and flexible frame structures (intra-frame period, GOP size). Moreover,
a simple filtering is added as a preprocessing option to target the low bitrate
track. In short, this paper illustrates how to obtain the best results
from a fully learned video codec in a practical context.

\section{The AIVC codec}

\begin{figure*}[ht]
  \centering
  \includegraphics[width=0.97\textwidth]{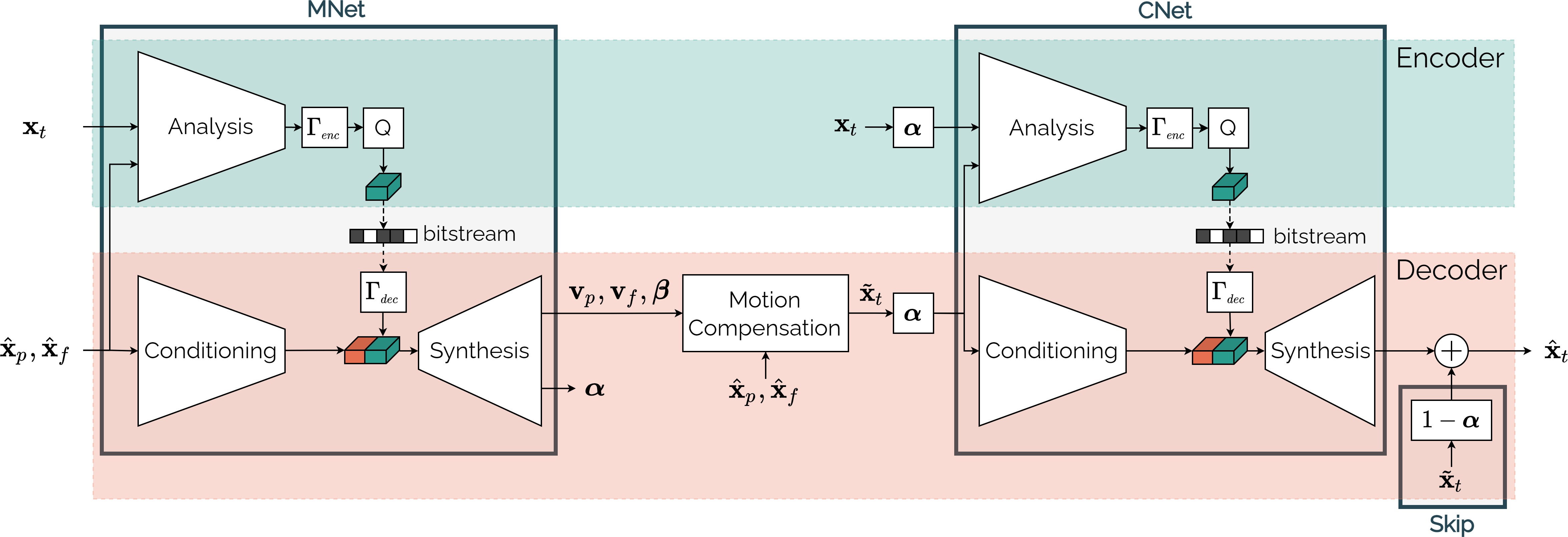}
  \caption{Block diagram of AIVC. $\Gamma_{enc}$ and $\Gamma_{dec}$ are
  feature-wise quantization gains as proposed in \cite{ladune2022aivc}.}
  \label{fig:global_diagram}
\end{figure*}

This section describes AIVC, an end-to-end learned video codec. More technical
details are provided in \cite{ladune2022aivc}, open-source models and several
illustrations are available at
{\small{\url{https://orange-opensource.github.io/AIVC/}}}.

\subsection{Architecture overview}

AIVC performs the coding of one video frame $\mathbf{x}_t$, while exploiting
information from up to two already transmitted (\textit{i.e.} available at the
decoder) reference frames $\pastref$ and $\futureref$. AIVC architecture is presented in
Fig. \ref{fig:global_diagram}. It is based on two \textit{conditional}
autoencoders:~MNet and CNet. Unlike the usual analysis-synthesis autoencoder, a
conditional autoencoder \cite{ladune2021conditional} features a decoder-side
conditioning transform. This conditioning transform aims to compute relevant
information from decoder-side quantities, which reduces the amount of bits sent
from the encoder to the decoder.
\bigskip

MNet stands for motion network and is responsible for estimating and conveying
motion information. This information is then used by a motion compensation
algorithm to obtain a temporal prediction $\pred$. Motion information is
composed of two pixel-wise motion maps $\pastflow$ and $\futureflow$ (one for
each reference frame) as well as a bi-directional prediction weighting
$\bbeta$. Furthermore, MNet also computes and conveys a coding mode selection
$\alpha$, which arbitrates between two possible coding modes.
\bigskip

The first coding mode is called Skip mode. It is a direct copy of the temporal
prediction, with no transmission associated. Consequently, Skip mode is
particularly dedicated to the stationary areas, well predicted enough to be
directly copied. The second coding mode relies on CNet to transmit the areas of
$\mathbf{x}_t$ not present in the prediction $\pred$ \textit{i.e.} the areas
for which the prediction is not accurate enough. In the end, both coding mode
contributions are summed up to obtain the decoded frame $\hat{\mathbf{x}}_t$.
\bigskip

The analysis and synthesis architecture proposed in \cite{cheng2020learned} is
used for MNet and CNet. The conditioning transform replicates the analysis
architecture. These transforms are based on convolutional layers with attention
modules and residual blocks. Moreover, AIVC relies on hyperpriors to perform the
entropy coding. The AIVC encoder has around 17 million parameters. The AIVC
decoder has 34 million parameters which requires 136 MBytes of storage.

\subsection{Training}

\begin{figure}[htb]
  \centering
  \includegraphics[width=0.5\linewidth]{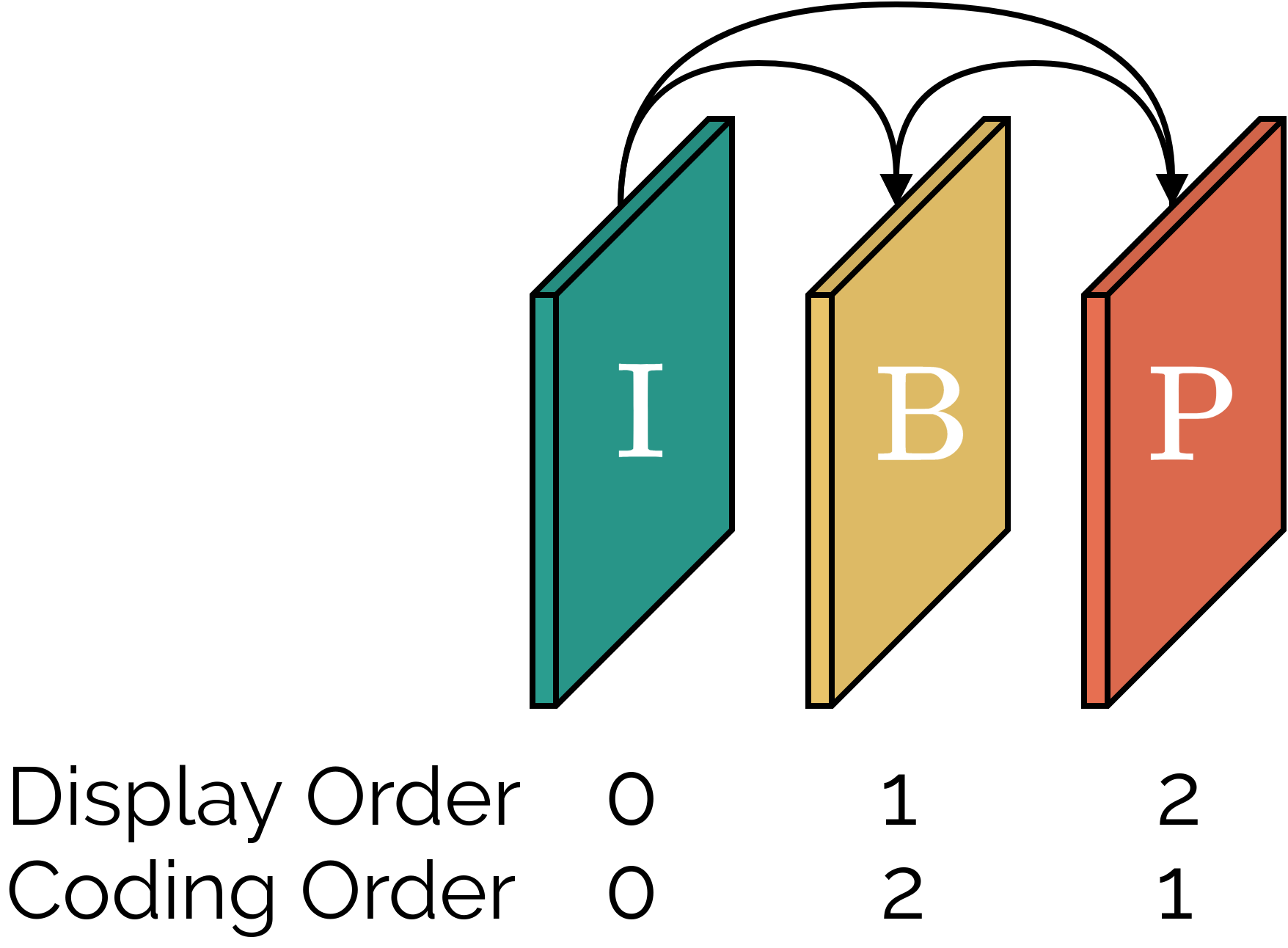}
  \caption{Training configuration.}
  \label{fig:training_config}
\end{figure}

The purpose of the training stage is to prepare the model to compress frames
with zero, one or two reference frames available. To this end, a training
example consists of the successive coding of 3 frames (see Fig.
\ref{fig:training_config}). This coding configuration features an I-frame (no
reference), a P-frame (one reference) and B-frame (two references). The model
parameters are learned through a stochastic gradient descent which aims to
minimize the following loss function over the 3 frames:
\begin{equation}
  \mathcal{L}_\lambda = \sum_{t=0}^{2} \mathrm{D}\left(\mathbf{x}_t, \hat{\mathbf{x}}_t\right)
  + \lambda \left(R_m + R_c\right).
\end{equation}

In the above equation, $\mathrm{D}$ denotes the distortion function between the
original and compressed frame. It is implemented based on the MS-SSIM. The
hyperparameter $\lambda$ balances the distortion with the sum of MNet rate $R_m$
and CNet rate $R_c$. The whole model is learned end-to-end from scratch and 8
different hyperparameters $\lambda$ are used to obtain 8 encoder-decoder pairs,
addressing various qualities.

\definecolor{lavenderpurple}{rgb}{0.59, 0.48, 0.71}
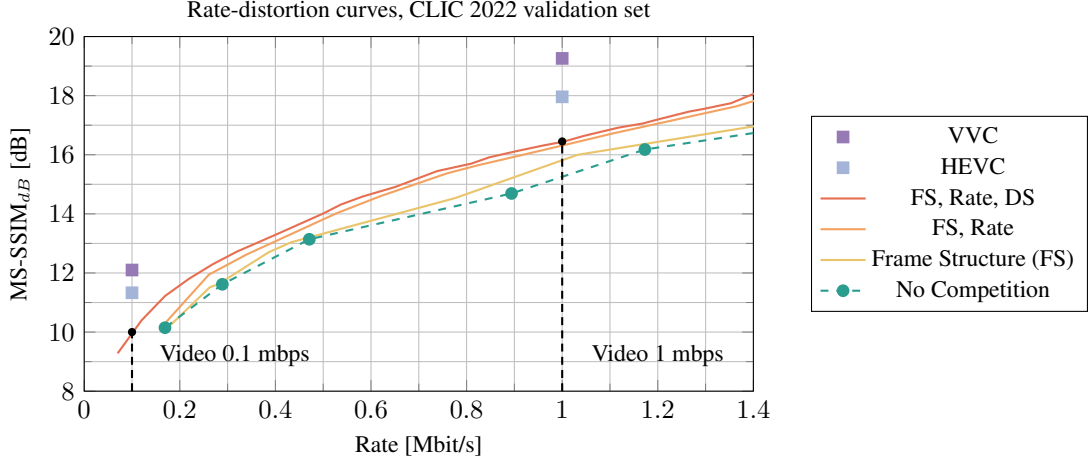
\begin{figure*}
  \centering
  \begin{tikzpicture}
      \begin{axis}[
          grid= both,
          xlabel = {\small Rate [Mbit/s]},
          ylabel = {\small $\text{MS-SSIM}_{dB}$ [dB]} ,
          xmin = 0, xmax = 1.4,
          ymin = 8, ymax = 20,
          ylabel near ticks,
          xlabel near ticks,
          height=6.3cm,
          width=0.6\linewidth,
          xtick distance={0.2},
          ytick distance={2},
          minor y tick num=1,
          minor x tick num=1,
          title={\small Rate-distortion curves, CLIC 2022 validation set},
          title style={yshift=-0.75ex},
          legend style={
              at={(1.5,0.5)},
              anchor=east,
          }
      ]

      \addplot[thick, lavenderpurple, mark=square*, only marks] coordinates {
        (0.1, 12.10)
        (1.0, 19.26)
      };
      \addlegendentry{\small VVC}

      \addplot[thick, lightblue, mark=square*, only marks] coordinates {
        (0.1, 11.33)
        (1.0, 17.96)
      };
      \addlegendentry{\small HEVC}

      \addplot[solid, thick, burntsienna, mark=no] table {data/gop_rate_ds_compete.txt};
      \addlegendentry{\small FS, Rate, DS}

      \addplot[solid, thick, sandybrown, mark=no] table {data/gop_rate_compete.txt};
      \addlegendentry{\small FS, Rate}

      \addplot[solid, thick, orangeyellow, mark=no] table {data/gop_compete.txt};
      \addlegendentry{\small Frame Structure (FS)}

      \addplot[dashed, thick, persiangreen, mark=*, mark options={solid}] table {data/no_compete.txt};
      \addlegendentry{\small No Competition}

      \draw [thick, densely dashed] (axis cs:1.0,8) -- (axis cs:1.0,16.4);
      \draw [thick, densely dashed] (axis cs:0.1,8) -- (axis cs:0.1,10);
      \draw [black,fill=black] (axis cs:1.0,16.45) circle (.3ex);
      \draw [black,fill=black] (axis cs:0.1,10) circle (.3ex);

      \node [align=left] at (axis cs:0.315,9.2) {\small Video 0.1 mbps};
      \node [align=left] at (axis cs:1.200,9.2) {\small Video 1 mbps};

      \end{axis}
  \end{tikzpicture}
  \caption{Benefits of per-sequence competition. FS stands for frame structure
  competition and DS denotes the possibility of downsampling. The black dots
  indicates the AIVC submissions to the challenge. Quality is measured as
  $\text{MS-SSIM}_{dB} = -10 \log_{10} \left(1 - \text{MS-SSIM}\right)$.}
  \label{fig:rd_results}
\end{figure*}

\subsection{Decoder conditioning}

During the encoding and decoding process, the latent variables of AIVC are
mapped to a binary code using the arithmetic coder \texttt{torchac}
\cite{mentzer2019practical}. This step is particularly sensitive to a potential
drift due to floating point computations. Such a drift might arise as the
encoding and decoding happen on two different devices. To circumvent this, the
decoder is made resilient to cross-platform encoding/decoding through a light
quantization of its internal parameters.

\section{Coding choices competition}
\label{sec:sequence_wise_competition}

\usetikzlibrary{decorations.pathreplacing}
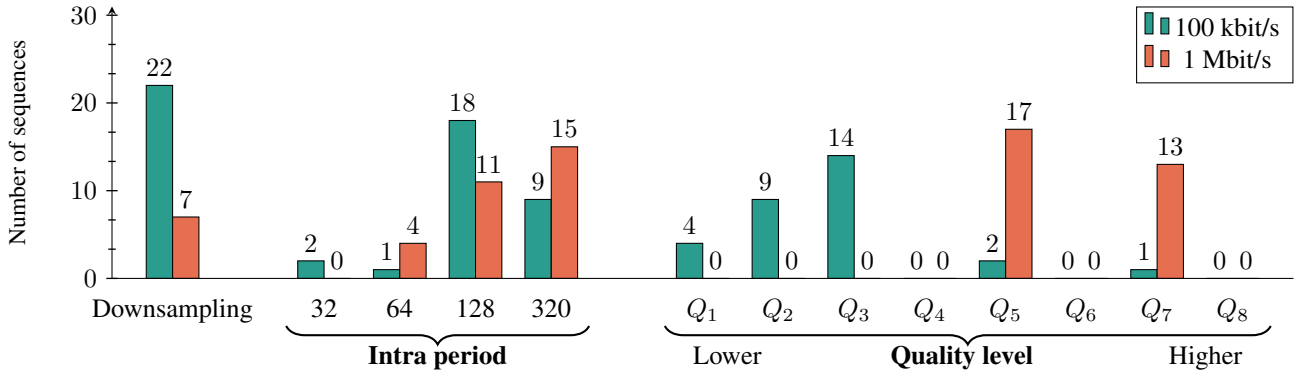
\begin{figure*}[htb]
  \begin{tikzpicture}
    \begin{axis}[
        height=5.2cm,
        width=0.99\linewidth,
        ylabel={\small Number of sequences},
        xmin=0.5, xmax=15.5,
        ymin=0, ymax=31,
        ytick distance=10,
        minor y tick num=2,
        enlarge x limits=0.02,
        xticklabels={,Downsampling,,32,64,128,320,,$Q_1$, $Q_2$, $Q_3$, $Q_4$, $Q_5$, $Q_6$, $Q_7$, $Q_8$},
        xtick={0,...,15},
        xtick pos=left,
        ytick pos=left,
        axis x line*=center,
        axis y line=left,
        xlabel style={right},
        bar width=0.35cm,                       
        ybar=0.0,                           
        xtick style={draw=none},
        ytick style={black, line width=0.5pt},
        axis on top,
        nodes near coords,
        every node near coord/.append style={
            /pgf/number format/.cd,
            fixed,
            fixed zerofill,
            precision=0
        },
        clip=false,
        legend entries={100~kbit/s,1~Mbit/s},
        legend style={at={(1,1)},anchor=north east},
        title style={yshift=-0.75ex},
        unbounded coords=jump,
      ]
    
    \addplot[ybar, fill=persiangreen] coordinates {
        (1,22)
        (3,2)
        (4,1)
        (5,18)
        (6,9)
        (8,4)
        (9,9)
        (10,14)
        (11,0)
        (12,2)
        (13,0)
        (14,1)
        (15,0)
    };

    \addplot[ybar, fill=burntsienna] coordinates {
        (1,7)
        (3,0)
        (4,4)
        (5,11)
        (6,15)
        (8,0)
        (9,0)
        (10,0)
        (11,0)
        (12,17)
        (13,0)
        (14,13)
        (15,0)
    };

    \draw [decorate,decoration={brace,amplitude=7pt,mirror,raise=4ex},thick]
    (axis cs:2.5,0) -- (axis cs:6.5,0) node[midway,yshift=-3em]{\textbf{Intra period}};

    \draw [decorate,decoration={brace,amplitude=7pt,mirror,raise=4ex},thick]
    (axis cs:7.5,0) -- (axis cs:15.5,0) node[midway,yshift=-3em]{Lower\hphantom{aaaaaaaaaaa}\textbf{Quality level}\hphantom{aaaaaaaaaaa} Higher};

\end{axis}
\end{tikzpicture}
\caption{Coding options selected for the CLIC 2022 validation set.}
\label{fig:codingchoices}
\end{figure*}

A per-sequence competition among the different coding choices of AIVC is
implemented to obtain better quality at both challenge rate targets. Given a set
of coding choices $\mathcal{C} = \left\{c_1, \ldots, c_N\right\}$,  AIVC
performs the coding of the video sequence with each coding choice $c_i$ to
obtain its rate-distortion cost:
\begin{equation}
  J_{c_i}\left(\lambda\right) = D_{c_i} + \lambda R_{c_i},
\end{equation}
Where $D$ denotes the distortion (based on the MS-SSIM), $R$ the rate and
$\lambda$ an external rate constraint. Finally, the optimal coding choice
$c^\star$ is chosen in order to minimize the rate-distortion cost of the video
sequence:
\begin{equation}
  c^\star = \underset{c_i \in\ \mathcal{C}}{\mathrm{argmin}}\ J_{c_i} \left(\lambda\right).
\end{equation}

\textbf{Frames structure}---The CLIC sequences exhibit various temporal
behaviors:~some sequences present important motions while others have large
motionless areas. In order to take into account this diversity, different frame
structures are tested by the AIVC encoder. The ability of AIVC to process I, P
and B frames is leveraged by evaluating several intra periods (distance between
I frames) and GOP sizes (distance between references). As such, an optimal frame
structure is chosen per-sequence. The green and yellow curves in Fig.
\ref{fig:rd_results} present the benefits of frame structure competition over
the initial models with a fixed frame structure.
\bigskip

\textbf{Rate allocation}---Some video sequences of the CLIC 2022 dataset can be
compressed into a smaller number of bits while still offering an acceptable
visual quality. As such, it is relevant to save bits for these easy-to-compress
sequences in order to invest those bits into more challenging sequences. In
practice, each sequence is compressed with the 8 encoder-decoder pairs and the
best one is chosen sequence-wise. The orange curve in Fig. \ref{fig:rd_results}
presents the benefits of adding the rate allocation competition to the frame
structure competition. This extended competition allows for a significant
increase in quality at the 1 Mbit/s rate target. Yet, the 100 kbit/s target
remains unreachable for our current systems.
\bigskip

\textbf{Preprocessing}---In order to compress videos at lower rates, an optional
preprocessing step is added to AIVC. This preprocessing consists of a bilinear
spatial downsampling of the video, which removes high-frequency content known
to be notably hard to compress. The video is then upsampled at the decoder-side
to recover the original resolution. The usage of the downsampling preprocessing
is added as a supplementary coding choice for the sequence-wise competition. The
red curve in Fig. \ref{fig:rd_results} presents the benefits of adding the
downsampling option to the frame structure and rate allocation competitions.
While this preprocessing option is added with the low-rate target in mind, it
turns out to be also slightly beneficial at higher rates.
\bigskip

In the end, performing per-sequence competition of the available coding modes
yields a significant performance increase compared to the raw AIVC models.
Indeed, using coding mode competition offers a BD-rate of $-26~\%$ \textit{i.e.}
it requires $26~\%$ less rate to achieve the same quality. This results in the
submissions to the challenge validation stage shown in Table
\ref{table:submissions}. The details of the different coding choices selected
for both submissions is presented in Fig. \ref{fig:codingchoices}. Compared to
the high-rate submission, the low-rate one more often uses the downsampling
option and relies on lower quality levels for the rate allocation.

\begin{table}[H]
  \centering
  \caption{Submissions for the validation stage.}
  \begin{tabular}{l|c c c c}
      \multirow{2}{*}{Name}  & Decoder size & Data size & PSNR  & Decoding\\
            & [MBytes]     & [MBytes]  & [dB]  & time [s] \\
      \midrule
      \midrule
      AIVC & 884.7 & 3.55 & 26.198 & 16~069 \\
      AIVC & 884.9 & 36.63 & 30.994 & 25~364 \\
  \end{tabular}
  \label{table:submissions}
\end{table}

Two modern video coding standards (HEVC and VVC) are used as anchors in Fig.
\ref{fig:rd_results}. For fair comparison, these coding standards also benefit
from rate allocation competition and also feature the downsampling option. While
our learned codec AIVC offers promising performance, conventional approaches
such as HEVC and VVC still obtain the best results. This may be due to their
extensive usage of competition, where each pixel of a video can be processed in
many different ways. This allows for fine-grain content adaptation and leads to
better coding performance.

\section{Limitations}

Even though AIVC results show that learned video coding is a promising field, a few
limitations still have to be overcome. The most promising area for improvement
is the introduction of more content adaptation mechanisms within the codec, as
discussed in Section \ref{sec:sequence_wise_competition}. In this section, two
others limitations of AIVC are discussed.
\bigskip

\textbf{Quality metric}---The issue of the quality metric used during the
training stage is yet to be solved. For convenience, this work relies solely on
the MS-SSIM to assess the quality of the decoded videos. However, recent work by
Mentzer et al. \cite{DBLP:journals/corr/abs-2107-12038} shows the inadequacy of
traditional quality metrics (PSNR, MS-SSIM, VMAF \cite{vmaf}, PIM \cite{pim},
LPIPS \cite{lpips}, FID \cite{fid}) to predict the actual (subjective) quality
of video content. The same work hints that an additional GAN-based distortion
yields significant quality improvements. A similar conclusion is reached by the
CLIC 2021 image track, where most of the top performing systems feature a
GAN-based distortion term. We believe that AIVC could obtain better perceptual
quality through the introduction of a GAN-based loss.
\bigskip

\textbf{Test-train mismatch}---A second limitation of AIVC is due to a
train-test mismatch. To achieve reasonable training time, AIVC is trained using
the smallest possible coding configuration which features a 2-frame intra
period. For the CLIC 2022 challenge, the shortest intra period used is 32 frames
and the longest is 320 frames. This causes an important change in the I frame
importance between the training stage and the test stage. Consequently, AIVC
learns a rate allocation strategy targeting a 2-frame intra period, which
largely deviates from the test stage. This inconsistency may have an impact on the
rate-distortion performance. This needs to be solved by using a bigger training
configuration, at the cost of a longer training time.

\section{Conclusion}

This paper presents the AIVC submission to the CLIC 2022 video track. AIVC is a
fully learned video codec, based on two conditional autoencoders. Thanks to its
flexibility in frame structure and rate allocation, AIVC allows the performing
of sequence-wise competition. Consequently, the optimal coding configuration is
used for each sequence. This results in a rate reduction of 26~\%.

By design, learned video codecs offer some interesting features for subjective
quality as they do not operate through block-based operations. Yet, learned
codecs are still outperformed by conventional video coding algorithms (HEVC,
VVC) for objective metrics such as MS-SSIM. We strongly believe that the
importance of competition in conventional systems allows for more content
adaptation, resulting in state-of-the-art performance. This hints at relevant
future works for learned coding, which should foster competition and content
adaptation through the introduction of additional coding modes.

{\small
\bibliographystyle{ieee_fullname}
\bibliography{refs}

\begin{thebibliography}{10}\itemsep=-1pt

\bibitem{hevc}
Gary~J. Sullivan, Jens-Rainer Ohm, Woo-Jin Han, and Thomas Wiegand.
\newblock Overview of the high efficiency video coding (hevc) standard.
\newblock {\em IEEE Transactions on Circuits and Systems for Video Technology},
  2012.

\bibitem{vvc}
Benjamin Bross, Jianle Chen, Jens-Rainer Ohm, Gary~J. Sullivan, and Ye-Kui
  Wang.
\newblock Developments in international video coding standardization after avc,
  with an overview of versatile video coding ({VVC}).
\newblock {\em Proceedings of the IEEE}, 2021.

\bibitem{cheng2020learned}
Zhengxue Cheng, Heming Sun, Masaru Takeuchi, and Jiro Katto.
\newblock Learned image compression with discretized gaussian mixture
  likelihoods and attention modules.
\newblock In {\em 2020 {IEEE/CVF} Conference on Computer Vision and Pattern
  Recognition}.

\bibitem{guo2021learning}
Zongyu Guo, Runsen Feng, Zhizheng Zhang, Xin Jin, and Zhibo Chen.
\newblock Learning cross-scale prediction for efficient neural video
  compression, 2021.

\bibitem{Agustsson_2020_CVPR}
Eirikur Agustsson, David Minnen, Nick Johnston, Johannes Balle, Sung~Jin Hwang,
  and George Toderici.
\newblock Scale-space flow for end-to-end optimized video compression.
\newblock In {\em Proceedings of the IEEE/CVF Conference on Computer Vision and
  Pattern Recognition}, 2020.

\bibitem{9506275}
David Alexandre, Hsueh-Ming Hang, Wen-Hsiao Peng, and Marek Domański.
\newblock Deep video compression for interframe coding.
\newblock In {\em 2021 IEEE International Conference on Image Processing
  (ICIP)}.

\bibitem{hu2020improving}
Zhihao Hu, Zhenghao Chen, Dong Xu, Guo Lu, Wanli Ouyang, and Shuhang Gu.
\newblock Improving deep video compression by resolution-adaptive flow coding.
\newblock In {\em Computer Vision - {ECCV} 2020 - 16th European Conference}.

\bibitem{hu2021fvc}
Zhihao Hu, Guo Lu, and Dong Xu.
\newblock {FVC:} {A} new framework towards deep video compression in feature
  space.
\newblock In {\em {IEEE} Conference on Computer Vision and Pattern Recognition,
  {CVPR} 2021}.

\bibitem{ladune2022aivc}
Théo Ladune and Pierrick Philippe.
\newblock {AIVC}: Artificial intelligence based video codec, 2022.

\bibitem{Philippe2021a}
P. Philippe and T. Ladune.
\newblock Coding standards as anchors for the {CVPR CLIC} video track.
\newblock In {\em 4th Challenge on Learned Image Compression}, Jun 2021.

\bibitem{ladune2021conditional}
Th{{\'{e}}}o Ladune, Pierrick Philippe, Wassim Hamidouche, Lu Zhang, and
  Olivier D{{\'{e}}}forges.
\newblock Conditional coding for flexible learned video compression.
\newblock In {\em Neural Compression: From Information Theory to Applications
  -- Workshop @ ICLR 2021}.

\bibitem{mentzer2019practical}
Fabian Mentzer, Eirikur Agustsson, Michael Tschannen, Radu Timofte, and Luc
  Van~Gool.
\newblock Practical full resolution learned lossless image compression.
\newblock In {\em Proceedings of the IEEE Conference on Computer Vision and
  Pattern Recognition (CVPR)}, 2019.

\bibitem{DBLP:journals/corr/abs-2107-12038}
Fabian Mentzer, Eirikur Agustsson, Johannes Ball{\'{e}}, David Minnen, Nick
  Johnston, and George Toderici.
\newblock Neural video compression using gans for detail synthesis and
  propagation.
\newblock 2021.

\bibitem{vmaf}
Netflix.
\newblock {VMAF} - video multi-method assessment fusion.
\newblock \url{https://github.com/Netflix/vmaf}.

\bibitem{pim}
Sangnie Bhardwaj, Ian Fischer, Johannes Ball{\'{e}}, and Troy~T. Chinen.
\newblock An unsupervised information-theoretic perceptual quality metric.
\newblock In {\em Annual Conference on Neural Information Processing Systems,
  NeurIPS 2020}.

\bibitem{lpips}
Richard Zhang, Phillip Isola, Alexei~A Efros, Eli Shechtman, and Oliver Wang.
\newblock The unreasonable effectiveness of deep features as a perceptual
  metric.
\newblock In {\em CVPR}, 2018.

\bibitem{fid}
Martin Heusel, Hubert Ramsauer, Thomas Unterthiner, Bernhard Nessler, and Sepp
  Hochreiter.
\newblock Gans trained by a two time-scale update rule converge to a local nash
  equilibrium.
\newblock In {\em Proceedings of the 31st International Conference on Neural
  Information Processing Systems}, 2017.

\end{thebibliography}
}

\end{document}